# Linear discriminant analysis based predator-prey analysis of hot electron effects on the *X*-pinch plasma produced *K*-shell Aluminum spectra

Mehmet Fatih Yilmaz[1,a], Yusuf Danisman[2], Jean Larour[3], Leonid Arantchouk[3]

[1] *Engineering Department, University of Dammam, Dammam, KSA*

[2] *Department of Mathematics, University of Oklahoma, OK, USA*

[3] *Laboratoire de Physique des Plasmas (LPP), Ecole Polytechnique, UPMC, CNRS, U-PSud, OBSPM, Palaiseau, France*

[a] Author to whom correspondence should be addressed. Electronic mail: fthyilmaz53@gmail.com

**Abstract**

Linear Discriminant Analysis (LDA) is applied to investigate the electron beam effects on the *X*-pinch produced *K*-shell Aluminum synthetic spectra. The radiating plasma is produced by the explosion of two 25-μm Al wires on a compact *L-C* (40kV, 200kA and 200ns) generator and time integrated spectra are recorded using de Broglie spectrographs. The electron temperature and density ($T_e$ = 80 eV and $n_e$ = 1x10$^{20}$ cm$^{-3}$), as well as the hot electron beam fraction ($f$ = 0.2 and energy centered at 10 keV) in the Al plasma, were extracted using coefficients of principal components derived from the non-*LTE K*-shell Al model. LDA shows that the weak transitions of Al Heα, Lyα and Heβ represents Stark splitting and shifting profiles. Finally, these transitions follows the elliptic polarization of Stokes V profiles in the presence of electron beam fraction. A 3-dimensional representation of LDA shows that the presence of electron beam, reflects outward spiral turbulence. These spirals show the signature of Langmuir turbulence. These spirals are modeled with logistic growth of predator-prey model. This modeling suggests that the electron beams and ions represent the preys and predator, respectively and the center region of the spirals tends to have low temperatures of 50-100 eV.

## I.  INTRODUCTION

Suprathermal electrons in plasmas are a hot topic in studies on inertial confinement fusion and high energy density physics. Hot electrons are diagnosed with many differents experimental and computational techniques: *x*-ray emission, electron breamsstrahlung and *K*α emission, spectropolarimetry and particle-in-cell modelling are some of them[1-4]. Collisional radiative models with non-Maxwellian electron distribution was an another alternative method to diagnose hot electrons in the emission spectra. Abdallah *et al.* compared synthetic and experimental emission spectra of laser produced *K*-shell Al plasma. The calculations employed an electron energy distribution which includes both thermal and hot electron components, as parts of a detailed collisional-radiative model. The comparison of spectra showed that the presence of hot electrons can alter the spectroscopic interpretation of electron density derived from standard thermal methods[5].

*X*-pinch produced plasmas are an alternative and unique source of hot electrons. Due to the existence of a strong electric field oriented along the axis of the interelectrode gap, it is expected that hot electrons affect collisional and radiative process in these plasmas. Moreover the very fast timescale, with subnanosecond *x*-ray bursts coming from hot spots, urges to use models far from the local thermodynamic equilibrium (*LTE*). Hansen *et al.* studied in detail that collisional excitation and ionization rates of *K*- and *L*-shell non-*LTE* collisional-radiative atomic kinetics models are highly sensitive to the fraction of hot electrons[6].

Modelling and relating individual line ratio changes in the spectra with hot electrons is a challenging work. Namely, we consider here the inclusion of new parameters in collisional radiative models, such as a fraction of hot electrons, meaning half-width half-maximum of electron beam energy standing beside plasma electron temperature. That is obviously increasing prosecution time and dimensions in the data pool. Since pattern recognition techniques such as principal component analysis (*PCA*) and linear discriminant analysis (*LDA*) can simplify a dataset into a lower dimension one, without great loss of information, they found many customers in spectroscopy of astrophysical plasmas[7-9].

Origin of the hot electrons generation, involved in the Langmuir turbulence, is another challenging work in high energy density plasmas[10]. There are many studies conducted on the complex relation between suprathermal electrons and Langmuir turbulence[11]. Recently, it has been shown that the interaction between electron beams and non-linear oscillations and micro-turbulence can be characterized with the predator-prey models[12-15].

In this work, we have applied linear discriminant analysis (*LDA*) method to investigate the electron beam effects on the spectra radiated by *X*-pinch produced, *K*-shell Al plasmas[16]. The *LDA* coefficients obtained are modelled by logistic growth with predator (predator-prey models) to investigate the relation between electron beams, ions and microturbulence[17]. The paper is organized as follows. After describing briefly the experiment in Sec. II, the third one gives the detail of the non-*LTE* collisional radiative model and studies the electron beam effects on the *K*-shell Al synthetic spectra applying linear discriminant analysis and predator-prey dynamics. Sec. IV discusses the modelling of experimental data and concluding remarks are given in Sec. V.

## II. EXPERIMENT

The mounting is a classical one and it has been described previously[16]. A compact pulsed power generator is devoted to the creation of point-like x-ray sources in the keV range for radiographical application to low contrast objects. The plasma is produced by the explosion of two 25-µm Al wires on a compact *L-C* (40kV, 200kA, 200ns) generator and the denser and brighter spots sit close to the crossing point, thus the device ensures creation of a localized and reproducible plasma. The time integrated spectrum is recorded on *x*-ray film through de Broglie *KAP, PET* or mica spectrographs installed in the equatorial plane. Due to the limited number of photons in the *K*-shell range, a rather wide (5 mm) entrance slit was used and it was not possible to get any spatial resolution along the *X*-pinch axis. Simultaneously, a set of fast detectors (filtered, absolute *XUV p-i-n* diodes and photoconductive diamonds) were recording the time dependence of *x*-ray flux in the keV region and in ns- and sub-ns regime (see Fig. 1). Time-integrated pinhole imaging in the same spectral region was performed radially on *DEF x*-ray film. The last two records were ensuring afterwards that there was a unique small-extension bright spot or, at least, a much brighter one, ensuring that there was no overlay of spectra coming from different sources nor geometrical blurring.

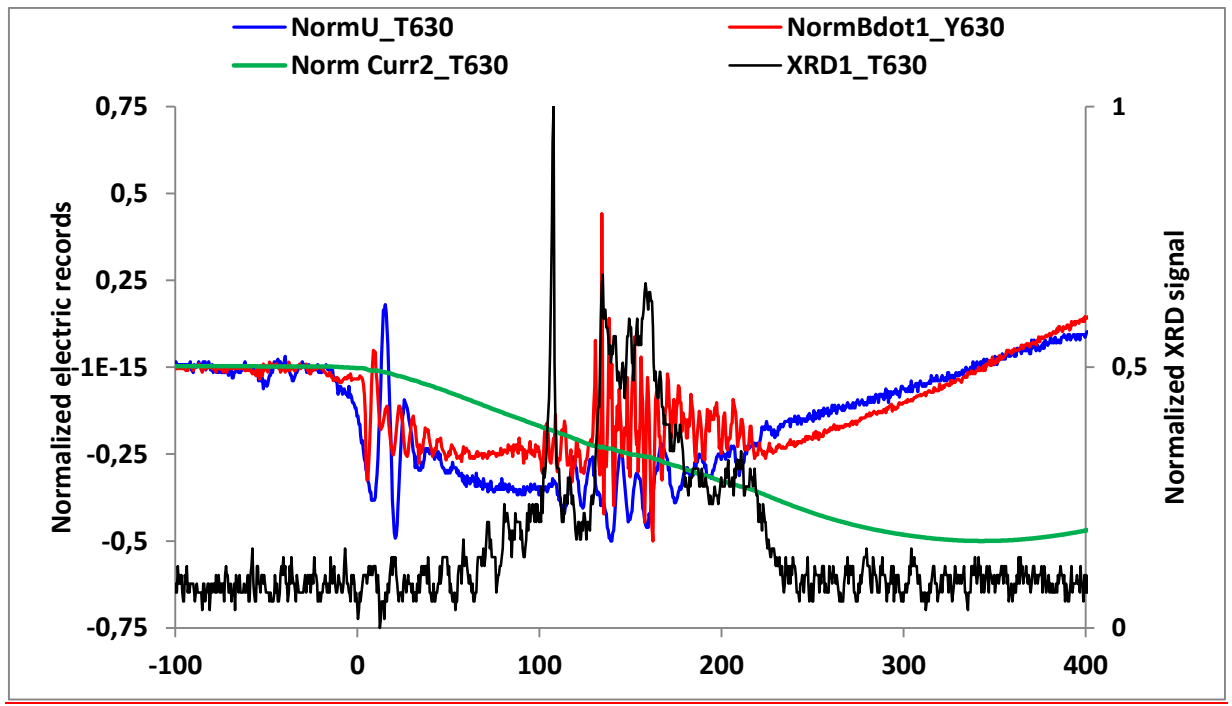

FIG 1. Close-up of the electrical and photonic records 500 ns around the time of pinching (shot *XP*_630). The electrical records (voltage, *B*-dot probe signal, current as numerically integrated from *B*-dot) are normalized (-1 to 1) to their span over the whole shot (left scale). *X*-ray signal is figured out by the *XRD* signal in volt (right scale), normalized (0 to 1). Close after the *x*-ray emission, a rise of the inductance leads to a current dip.

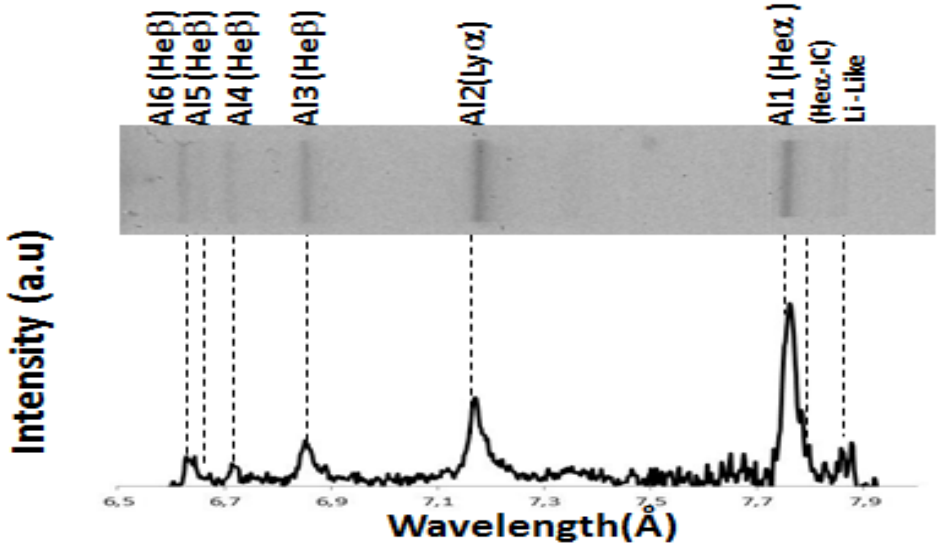

FIG.2 Time integrated spectrum of *K*-shell Al plasma (shot XP_630) with diagnostically important lines.

Over many tests with various wires, showing evidences of *K*-shell Al, *L*-shell Cu and Mo, the present work deals with the electron temperature and density and hot electron beam fractions of *X*-pinch produced *K*-shell Al plasma. Figure 2 presents a time-integrated *x*-ray spectrum of *K*-shell Al plasma (shot XP_630). The shape and intensity of lines are well resolved in the resonant transition of Al$_1$ (He-like), Al$_2$ (H-like), Al$_3$ (He-like) and Li-like Al as well as satellite transitions of Al$_1$, Al$_2$ and Al$_3$ (see[16]).

## III. ELECTRON BEAM EFFECTS ON THE SYNTHETİC SPECTRA OF *K*-SHELL ALUMINUM

### III.a. NON-LTE COLLISIONAL RADIATIVE MODEL OF *K*-SHELL AL

The details of collisional radiative model of *K*-shell Al and the principal component analysis have been described previously[16]. In brief, the energy level structures, spontaneous and collisional rates, collisional and photoionization cross-sections calculations were performed using the Flexible Atomic Code (FAC)[18]. Electron distribution function which was used in this model $F(e) = (1-f)*F_m + f*F_{nm}$ includes both Maxwellian ($F_m$) and non-Maxwellian ($F_{nm}$) distributions. The fraction of hot electrons was described by a Gaussian distribution centered at the characteristic energy $E_0 = 10$ keV, in order to be able to see more variations in resonance lines (H-like and He-like Al), rather than satellite lines of *K*-shell Al. Voigt profiles with the resolution of 300 were used to fit the line broadening of the experimental spectra[16].

## III.b. LINEAR DISCRIMINANT ANALYSIS

The effects of electron beams in the emission spectra can be diagnosed by the ratio of the lines which are sensitive to the high energy electron population, such as the ratio of different ion charge states of (Al1+Al1-IC)/Al2 and same ion charge states of (Al1+Al1-IC)/Al3[16].

One can also use pattern recognition techniques as an alternative plasma diagnostic. Principal Component Analysis (*PCA*) and Linear Discriminant Analysis (*LDA*) are the most common multivariable techniques to analyze the structure of a large set of data. Recently, *PCA* has been applied to diagnose electron beam effects in *X*-pinch produced *K*-shell Al plasmas and the details of the extracted principal components over *K*-shell Al model can be found elsewhere[16]. However, the brief explanation of *PCA*, *LDA* and unified version of *PCA* and *LDA* are given as below.

*PCA* is an unsupervised method which reduces the dimension of the data but preserves its characteristics. In *PCA*, the vectors which have the largest variance associated to the data, are computed and used as a basis for the new reduced space. Therefore, each original data is represented by its coordinate vector in the reduced space with a lower dimension. In the present paper, a 456-dimension space of original spectra will be projected onto a space of 3 dimensions. Therefore, original spectra can be visualised in a 3-dimension vector space.

Let $\Gamma$ be a matrix of size $N \times M$, the columns of which consist of the original dataset. Let $\mu$ be the mean value of the columns of $\Gamma$, and $\Phi$ be the matrix obtained from $\Gamma$ by subtracting $\mu$ from the each column of $\Gamma$. The covariance matrix, which is a measure of how much variables change together, is $\frac{1}{M}\sum_{i=1}^{M} \Phi_i \Phi_i^t$ where $\Phi_i$ is the *i*-th column of $\Phi$ and superscript *t* means transposed. The vectors which has largest variance are the eigenvectors of covariance matrix with largest eigenvalues, called principal components and denoted by |PC1>, |PC2>, |PC3>, … depending on the order of their eigenvalues.

In applications, to reduce the dimension, the dataset is projected onto the space spanned by the principal components which correspond to the largest eigenvalues. For instance, let |PC1>, |PC2>, |PC3> be the principal components which corresponds to the largest three eigenvalues. If an element *v* of the dataset is projected onto the space spanned by |PC1>, |PC2> and |PC3>, the projection vector is $\sum_{i=1}^{3}(v \bullet |PCi>) |PCi>$. Therefore in the new three dimensional space, the coordinate of *v* is (*v*●|PC1>, *v*●|PC2>, *v*●|PC3>), where ● is the dot product in real numbers.

In contrary, *LDA* is a supervised method which also reduces the dimension of the dataset. However, in *PCA* data is considered in its entirety whereas in *LDA* the focus is on the characteristics of the different classes to discriminate them. Let a dataset consisting of *K* classes of $N \times 1$ vectors, where each class contains $M$ vectors. Let $\Gamma_i^j$ be the i'th element of the class $j$ for $i = 1, 2, ..., M$ and $j = 1, 2, ..., K$. and $\mu_j$ be the mean of the class $j$ and $\mu$ be the mean of all classes. Then the within-class scatter matrix $S_w$ and the between-class scatter matrix $S_b$ can be expressed respectively :

$$S_w = \Sigma_{j=1}^{K} \Sigma_{i=1}^{M} (\Gamma_i^j - \mu_j)(\Gamma_i^j - \mu_j)^t \qquad (1)$$

$$S_b = \Sigma_{j=1}^{K} (\mu_j - \mu)(\mu_j - \mu)^t. \qquad (2)$$

In *LDA*, the eigenvectors of $(S_w)^{-1}.S_b$, which correspond to the largest eigenvalues, are considered and they are denoted by |LD1>, |LD2> and |LD3> for the first three of them. In *LDA*, there is a difficulty of taking the inverse of the matrix $S_w$ if it is large. Therefore *LDA* has difficulty in processing high dimensional data. To remedy this problem, in applications before performing *LDA* algorithm, dimension of the data is reduced by *PCA*. In this paper, a unified *PCA* and *LDA* algorithm is applied to the spectra clustering[19-22].

### III.c. SPECTRAL REPRESENTATION OF *LD* VECTORS

In the present work, as a first step, *PCA* is applied to the data obtained for different electron beam fractions separately. For each fraction, five densities are considered: $1 \times 10^{19}$, $5 \times 10^{19}$, $1 \times 10^{20}$, $5 \times 10^{20}$ and $1 \times 10^{21}$ cm$^{-3}$. Each density consists of a spectrum for the temperatures of 50, 60, 70, ... 500 eV (*i.e.* 46 different temperatures). Hence, in total, $5 \times 46 = 230$ spectra of size $456 \times 1$ are obtained. By applying *PCA*, the dimension is reduced to 40 by projecting each of the spectra into the space spanned by the most dominant 40 principal components (|PC1>, |PC2>, ..., |PC40>). As a second step, *LDA* is applied to the 230 spectra of size $40 \times 1$. For each spectrum, an |LD1> coefficient is obtained by projecting it onto the space spanned by |LD1>. Hence, in total, 230 |LD1> coefficients are obtained. |LD2> and |LD3> are computed as well.

In Fig. 3, the spectra of |LD1> are given for the beam fraction of $f = 0.00$ up to 0.20. Figure 3 shows that resonant transitions of Al1, Al2, Al3, Al4, Al5 and Al6 are almost absent but their intercombinations and dielectronic satellites are present. Figure 3.a shows that intercombinations and dielectronic satellites have splitting and shifting profiles. This effect is expected due to harmonic Stark effects and addition of fraction of electron beams in Fig. 3.b polarize these lines in elliptical shape of Stokes V profiles[23,24] Comparing the response of resonant transitions to the electron beams in *PCA* vector spectra, in our recent works, shows that *LDA* can clearly discriminate the behavior of weak transitions in the presence of electron beams[16].

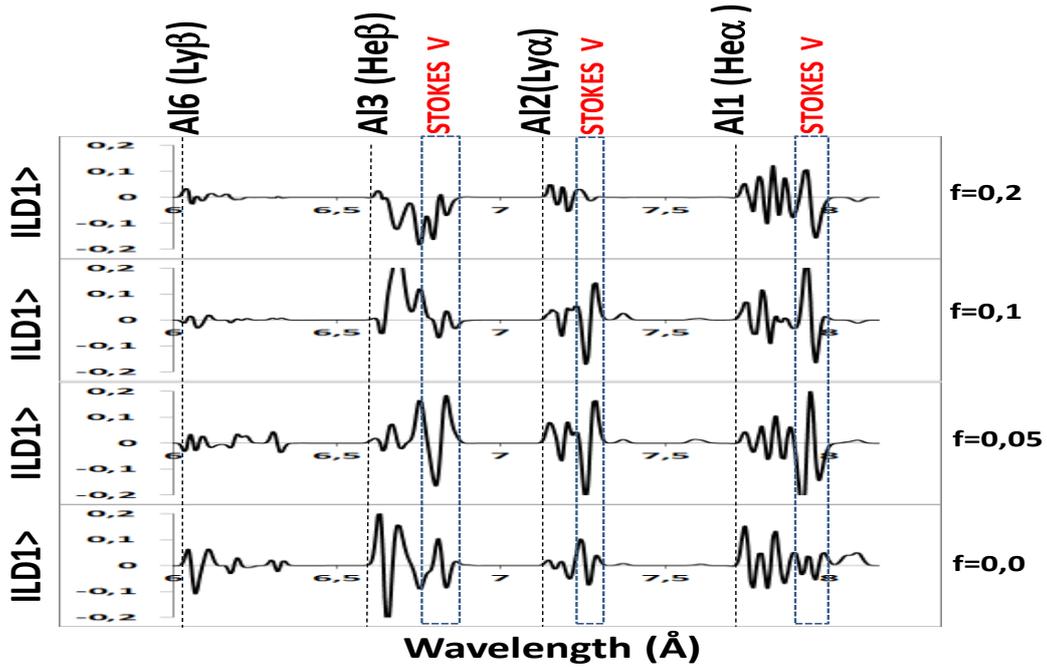

FIG. 3. Main features of the |LD1> spectra, (a) without electron beam fraction, $f = 0.00$ and (b) with beam fractions, $f = 0.05$, 0.10 and 0.20.

### III.d. THREE-DIMENSION REPRESENTATION OF LDA AND PREDATOR-PREY DYNAMICS

Figure 4 shows the behavior of |LD1>, |LD2> and |LD3> coefficients with temperature increase for the fraction group of $f = 0.00$, 0.05, 0.10, 0.15 and 0.20. The fraction case especially shows that the temperature increase results satellite transitions to move in an outward spiral turbulence.

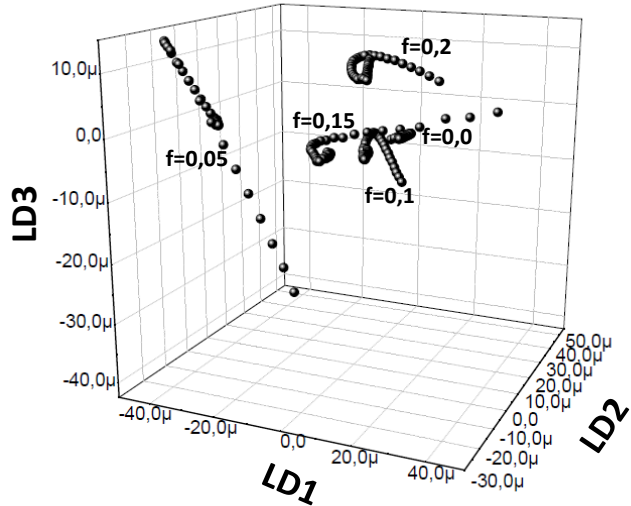

FIG. 4. |LD1>, |LD2> and |LD3> coordinates for different electron beam fractions ($f$ = 0.00, 0.05, 0.10, 0.15 and 0.20) at electron density of $n_e$ = 1x10$^{20}$ cm$^{-3}$. The μ value has been introduced in Sec.III.b.

**Predator-Prey Models**

Gurcan et al. stated the turbulence in hot dense plasma can be characterized by the predator-prey of population models[14,15]. The populations of two species interacting as a predator and prey can be modeled by using a pair of nonlinear, first-order equations which are the modified versions of the original Lotka–Volterra equations[26]. These models are called Predator-Prey models and they have been applied to many different areas such as chemical reactions[33], astronomy[34], economics[28-31], plasmas[13,15,22], and evolutionary game theory[34], to express the complicated real life situations as differential equations.

The Lotka–Volterra equations[26] are two non-linear and first-order differential equations:

$$\frac{dP}{dt} = [r - sQ]P \quad \text{and} \quad \frac{dQ}{dt} = (-u + vP)Q \quad (3)$$

where $P$ and $Q$ are populations of prey and predator, respectively and $r, s, u, v$ are parameters describing the interaction of the two species. Originally, these equations were used to exhibit the relation in a biological system of two interacting species: predator (Q) and prey (P).

Introducing a constraint on carrying capacity of the prey population forces to modify the original Lotka–Volterra equations and yields[25]:

$$\frac{dP}{dt} = [r\left(1 - \frac{P}{K}\right) - sQ]P \quad \text{and} \quad \frac{dQ}{dt} = (-u + vP)Q \quad (4)$$

where $K$ is the maximum size of preys. This is modified Lotka-Volterra model with capacity constraint. For a particular choice of parameters the graph of P versus Q is given in Fig 5.

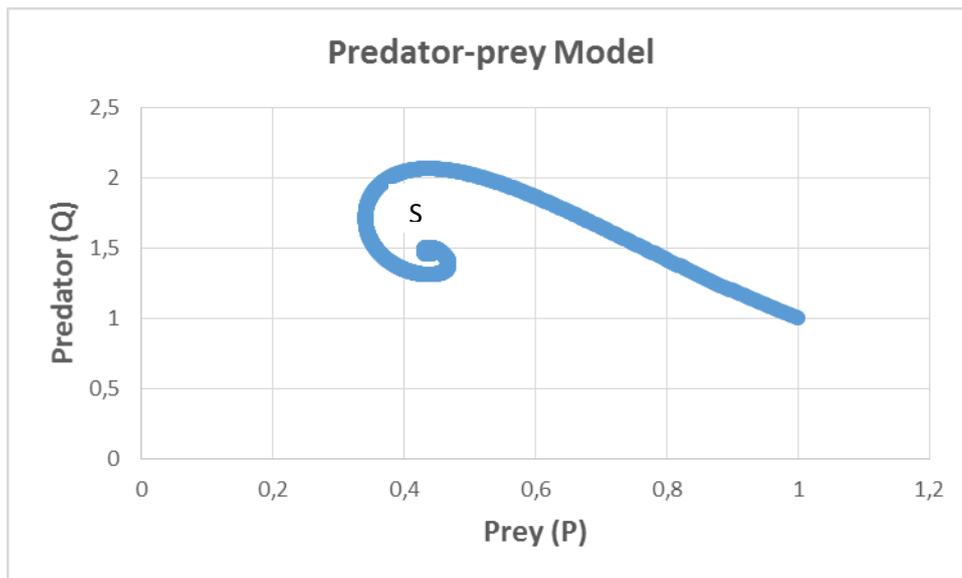

FIG 5: Scheme of a system trajectory in Predator-Prey model.

There is a steady state at the point $S = [\frac{u}{v}, \frac{r}{s}(1 - \frac{u}{vK})]$ and trajectory spirals around the steady state point *S*. There are two cases, either trajectories will spiral inwards and points converge to *S* or outwards and the points diverge from *S*.

In our case, Predator-Prey model is used to exhibit the characteristics of the plasma via linear discriminant analysis coefficients in Fig. 3. As a result of linear discriminant analysis for *f* = 0.2, it is shown that |LD2> coefficients behave like prey and |LD1> coefficients behave like predator as in Fig. 2, because |LD1> coefficients deplete the |LD2> coefficients. As it is known that radiative and dielectronic recombination transitions associated to the resonant transtions are due to electron capturing process, |LD2> and |LD1> coefficients represent the ions and electrons respectively[13]. Since the trajectory spiral outwards, the steady state point *S* is unstable. In contrary, in the case of *f* = 0.0 the trajectory spiral inwards and the central point is stable.

An important result of our work is, the point *S* corresponds to the lowest temperature and the center region of the spiral has lower temperatures (50-100 eV). On the other hand, on these low temperatures, coefficients are more stable in which the low ionization is fixed by the electron beams[13]. Therefore, as the temperature decreases (|LD2> coefficient, |LD1> coefficient) points converge to *S* which is an accumulation point. The temperature path is given in Fig. 6 and, as temperature decreases to 300 eV, |LD1> and |LD2> coefficients have inverse correlation. Above this temperature |LD1> and |LD2> coefficients are accumulated.

Another key finding is, by using the Predator-Prey model, the trend of the |LD1>-|LD2> coefficients can be predicted. This estimation is more accurate than the ones in[16,32] because in these papers only the information obtained by |PC1> is used whereas in this work |LD1> and |LD2> are together, which means that less information is lost, thus used for characterization.

Since the Predator-Prey model that we are using has a carrying capacity for the |LD2> coefficients, we are also able to find a boundary for the |LD2> coefficients of the plasma.

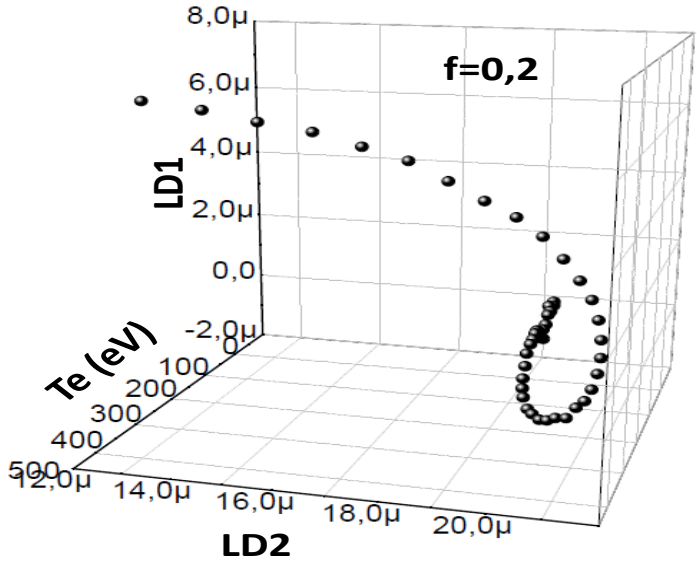

FIG. 6 |LD1⟩ and |LD2⟩ coefficient *vs* plasma electron temperature.

## IV. MODELLING USING *LDA*, *PCA* AND NON-*LTE* MODELS

In Fig. 7, for the electron densities of $n_e$ = 1x10$^{19}$, 1x10$^{20}$ and 1x10$^{21}$ cm$^{-3}$, the |LD1⟩ coefficients of the 46x3 spectra, in the temperature range between 50 and 500 eV, are given for the beam fractions of $f$ = 0.00, 0.01 and 0.20. Figure 7 shows that |LD1⟩ coefficients are more stabilized at $f$ = 0.1 at plasma electron density of 1x10$^{19}$ cm$^{-3}$. For the electron density of 1x10$^{20}$ cm$^{-3}$, |LD1⟩ coefficients follow a linear motion for the electron temperatures below 250 eV and non-linear motion above that temperature. For the electron density of 1x10$^{21}$ cm$^{-3}$, |LD1⟩ coefficients follow a nonlinear profiles for the electron temperatures below 250 eV and an almost linear motion above this temperature.

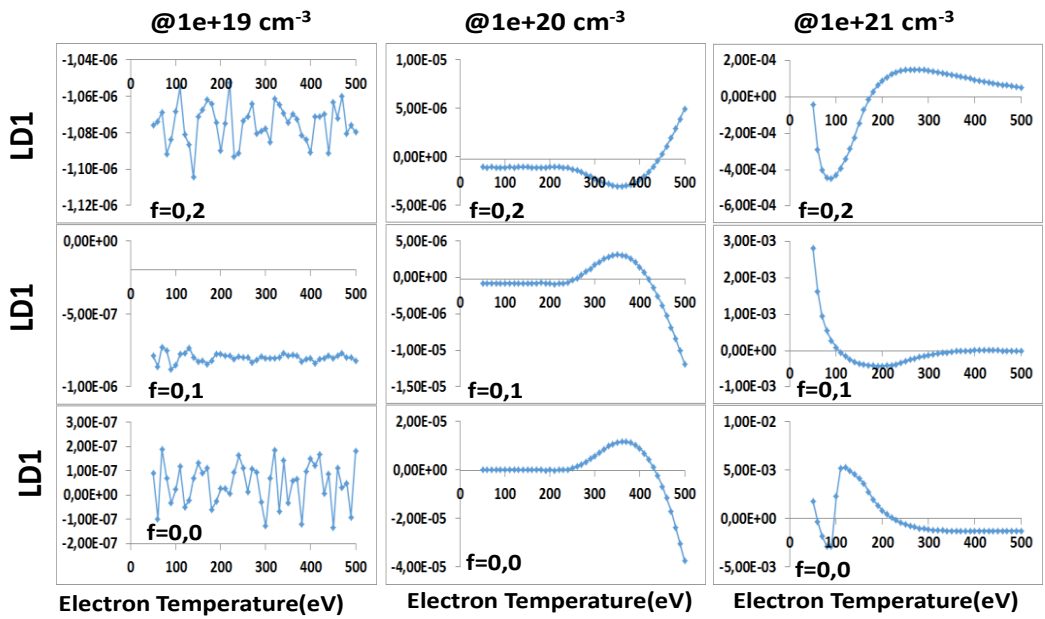

FIG. 7. Correspondence of |LD1>coefficients and electron temperatures at classified electron densities and beam fractions.

When it is compared to the traditional ratio diagnostics of *K*-shell Al spectra, there is no resemblance to both of (Al1+Al1-IC)/Al2 and (Al1+Al1-IC)/Al3 ratios. This result is expected because *LDA* focuses on the behavior of the weak transitions rather than strong characters of resonant transitions in *PCA* modeling[11].

The electron temperature of test for experimental spectrum can be estimated using the regression curve of these classes. The electron temperature of shot *XP*_630 has been fitted by a third degree polynomial using *LD* coefficients. This modelling gives the plasma electron temperature of 80 eV and density of $n_e = 1 \times 10^{20}$ cm$^{-3}$ and $f = 0.20$. *LDA* modelling overestimates the Al1 and Al2; since *LDA* is orthogonal to the *PCA* and mainly focuses transitions from upper levels of *y* and *z* coordinates, this result is expected. *PCA* and non-*LTE* modelling estimates the resonant transtions well and *PCA* modelling gives plasma electron temperature of 80 eV, density of $n_e = 1 \times 10^{20}$ cm$^{-3}$ and $f = 0.2$ (Fig. 8.b). Non-*LTE* modelling gives plasma electron temperature of 80 eV, density of $n_e = 1 \times 10^{20}$ cm$^{-3}$ and $f = 0.20$ (Fig. 8.c).

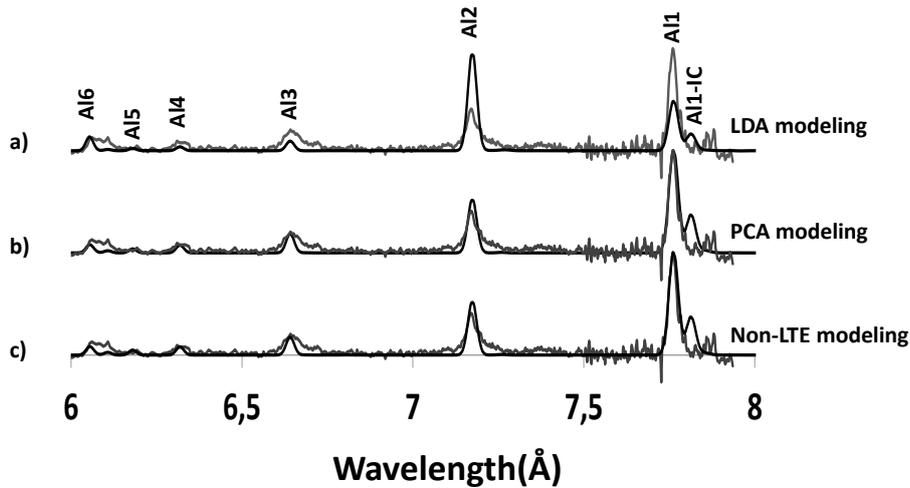

FIG. 8. Comparison of experimental spectrum with *LDA*, *PCA* and non-*LTE* model produced synthetic spectrum ($T_e = 80$ eV, $n_e = 1 \times 10^{20}$ cm$^{-3}$ and $f = 0.2$).

## V. CONCLUSION

As a first result of the present work, the *LDA* can be used for the data classification of non-*LTE* collisional radiative *K*-shell Al model and each spectrum can be characterized by the dominant *LD* coefficients. The *LDA* can also be used as an alternative plasma diagnostic of *K*-shell Al spectra, especially for lines with weak transitions of radiative and dielectronic satellites. However, *PCA* realizes a better correspondence with line ratio diagnostics. Modelling of a representative *K*-shell Al spectrum, using *LD* coefficients, gives $T_e$ = 80 eV, $n_e$ = 1x10$^{20}$ cm$^3$ and $f$ = 0.2 (with the beam energy centered at 10 keV). The *LDA* vector spectra show that addition of the electron beam leads the weak transitions of Al1, Al2 and Al3 to move in Stark splitting and shifting profiles. Addition of electron beam fractions proceed the weak transtions to have an elliptical polarization profiles. The plot of electron temperature, |LD1>, |LD2> and |LD3> coefficients (at electron density of $n_e$ = 1 x 10$^{20}$ cm$^{-3}$) clearly shows that electron beam addition on the spectral model generates quantized clusters in the vector space and move the weak transitions in a outward spiral like turbulence[18]. Modelling of these turbulence using Predator-Prey model suggests that center of the spiral have lower electron temperature in which ionization is fixed by the electron beams. Another result of Predator-Prey modeling is that ions and electrons behave as the predator and prey, respectively.